\documentclass[11pt]{article}

\usepackage[parfill]{parskip}
\usepackage{setspace}
\usepackage[margin=1in]{geometry}

\usepackage{amsmath}
\usepackage{amssymb}
\usepackage{graphicx}
\usepackage{array}
\usepackage{adjustbox}
\usepackage[percent]{overpic}
\usepackage{float}
\usepackage{upgreek}
\usepackage{bm}

\usepackage[round,authoryear]{natbib}
\usepackage{xcolor}
\usepackage{hyperref}
\hypersetup{colorlinks = true, allcolors = blue}
\usepackage{authblk}
\usepackage{booktabs}

\newcommand{\keywords}[1]{\par\medskip\noindent\textbf{Keywords:} #1}

\title{Bridging powder and multi-crystal diffraction with basis-adaptive texture tomography}

\author[a]{Martin Sæbye Carøe (msaca@dtu.dk)}%

\author[c]{Mads Allerup Carlsen}%

\author[d]{Felix Tristan Frankus}%

\author[b]{Adam André William Cretton}%

\author[d]{Michela La Bella}%

\author[b,e]{Innokentiy Kantor}%

\author[f,e]{Mads Ry Vogel Jørgensen}%

\author[b]{Henning Friis Poulsen}%

\author[a]{Jakob Sauer Jørgensen}%

\author[d]{Nils Axel Henningsson}

\affil[a]{Department of Applied Mathematics \& Computer Science, Technical University of Denmark, 2800 Kgs. Lyngby, Denmark}

\affil[b]{Department of Physics, Technical University of Denmark, 2800 Kgs. Lyngby, Denmark}

\affil[c]{Paul Scherrer Institut, 5232 Villigen PSI, Switzerland}

\affil[d]{Department of Civil and Mechanical Engineering, Technical University of Denmark, 2800 Kgs. Lyngby, Denmark}

\affil[e]{MAX IV Laboratory, Lund University, Fotongatan 2, 225 94, Lund, Sweden}

\affil[f]{Department of Chemistry \& iNANO, Aarhus University, Langelandsgade 140, Aarhus C., 8000, Denmark}

\begin{document} 
\maketitle

\begin{abstract}
In spatially resolved X-ray diffraction experiments using narrow beams, diffraction patterns from polycrystalline materials often fall between two well-served limits. Fine-grained, weakly textured microstructures produce smooth Debye--Scherrer rings suited to powder- and tensor-tomography methods, whereas coarse, weakly deformed grains produce isolated spots that can be indexed grain by grain. Many experimentally important polycrystalline materials, including plastically deformed metals, martensitic and ferroelastic materials containing complex twin microstructures, and geological aggregates with strong texture or heterogeneous grain size, produce spotty diffraction rings with broadened and overlapping peaks that fall between these limits. Texture tomography addresses this intermediate regime by reconstructing spatially resolved orientation distributions from diffraction data. In this intermediate regime, conventional texture tomography lacks the angular resolution needed for sharp orientation distributions, while grain-by-grain indexing can introduce grain boundary artifacts and underestimate intragranular misorientation. This work introduces basis-adaptive texture tomography. Candidate orientations obtained from peak indexing are used to replace the uniform orientation grid by a data-driven sparse basis set, thus achieving higher angular resolution. Like conventional texture tomography, the method benefits from reconstructing a full orientation distribution function in each voxel, allowing voxels to retain contributions from multiple grains, subgrains or domains rather than being forced into a single orientation. Simulated aluminum polycrystals show improved delineation of grain and sub-grain boundaries accompanied by lower intragranular orientation errors compared with uniform-basis texture tomography and point-by-point scanning 3DXRD. An experimental demonstration on tensile-deformed aluminum shows that bulk grain and subgrain structures with orientation spreads of several degrees can be mapped.
\end{abstract}


\section{Introduction}

Polycrystalline materials are central to engineering and geoscience applications, including structural alloys, additively manufactured metals, and geological aggregates. Their macroscopic properties are influenced by the crystallographic microstructure, including grain morphology, crystallographic texture, and elastic strain distributions. These properties influence mechanical strength, plastic deformation and crack initiation and propagation. Consequently, characterizing these quantities is essential for understanding and optimizing material performance.

Surface-sensitive techniques such as electron backscatter diffraction (EBSD) provide orientation mapping in high spatial resolution and are widely used for crystallographic texture analysis \citep{brent1993, humphreys2001, wilkinson1996, schwartz2009}. 
EBSD can be extended to volumetric mapping by advanced serial sectioning methods \citep{xu2007, zaefferer2009, echlin2015}.  

 For \emph{non-destructive} mapping of the crystallographic microstructure of bulk polycrystalline materials X-ray diffraction contrast imaging methods are prevalent. These provide access to embedded crystal orientation and strain states as spatially resolved fields. Existing techniques include three-dimensional X-ray diffraction (3DXRD) and related high-energy diffraction methods \citep{poulsen2001, poulsen2012, reischig2020, Bernier2020}, scanning 3DXRD \citep{hayashi2015, kim2023, henningsson2024}, diffraction contrast tomography \citep{johnson2008}, dark-field X-ray microscopy \citep{simons2015, Poulsen2020}, differential-aperture and Laue microdiffraction \citep{chung1999, larson2002}, and X-ray diffraction computed tomography (XRD-CT) \citep{harding1987, Bleuet2008}.

\begin{figure}[tbp] %
\begin{center}
\includegraphics[width=\textwidth]{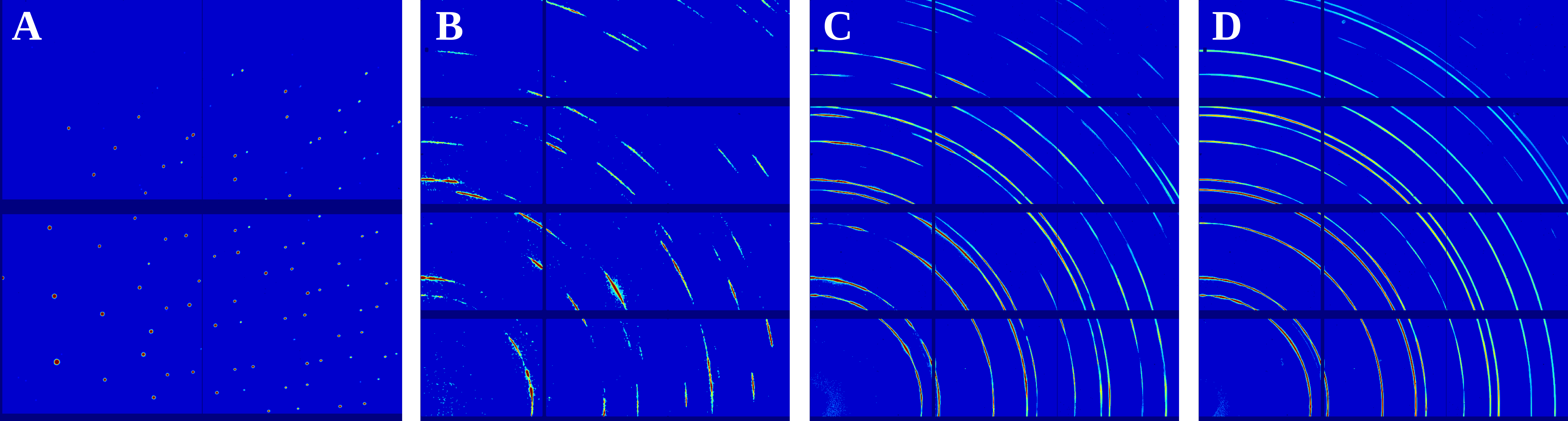} % NB use pdflatex for non-postscript
\end{center}
\caption{The spectrum of diffraction regimes. (A):~Diffraction pattern of a single crystal rotated $30^\circ$ (discrete diffraction regime). (D):~Isotropic diffraction pattern. (B) and (C):~Diffraction patterns in the intermediate regime stemming from sharp and weak textures respectively.}
\label{regime_plot}
\end{figure}

In this paper, we shall focus on the use of a monochromatic incoherent pencil beam. In this case the measured diffraction signal spans a spectrum from continuous to spotty depending on the size of the beam and the underlying microstructure (Figure~\ref{regime_plot}). At one extreme lies the \emph{isotropic diffraction regime}, where the beam illuminates many randomly oriented crystallites which are much smaller than the beam. Their contributions to the diffraction signal superpose to form smooth, continuous Debye--Scherrer rings (Figure~\ref{regime_plot}D). At the opposite extreme lies the \emph{discrete diffraction regime}, where the illuminated volume comprises a small number of undeformed or weakly deformed grains, each producing a set of isolated diffraction peaks that can be individually resolved on the detector (Figure~\ref{regime_plot}A).

Reconstructing the microstructure in these two regimes is addressed by fundamentally different classes of algorithms. In the isotropic diffraction regime, powder-based approaches such as X-ray diffraction computed tomography (XRD-CT) combined with Rietveld refinement \citep{rietveld1969} use azimuthally integrated ring intensities to recover spatially resolved lattice and phase information \citep{Bleuet2008}. For this algorithm, sharp peaks on the detector images result in artifacts \citep{vamvakeros2015}. In the discrete diffraction regime, algorithms associated with scanning 3D X-ray diffraction microscopy (s3DXRD) \citep{hayashi2015, Hayashi2019, Henningsson2020, Li2023} segment and index individual diffraction peaks to reconstruct grain- or domain-resolved orientation and strain. However, these algorithms are limited by the complexity of the microstructure: in case of highly deformed materials, strong textures or a large number of crystallites within the gauge volume they may fail. 

Between these two extremes, polycrystalline samples exhibit varying degrees of texture, and the diffraction pattern transitions from smooth to spotty rings. Texture tomography is a symmetry-informed extension of tensor tomography \citep{liebi2015, malecki2014, carlsen2024}, enabling reconstruction of the spatially varying orientation distribution function (ODF) from azimuthally varying Debye--Scherrer ring intensities \citep{frewein2024, carlsen2025a, carlsen2025c}. Because it models the full azimuthal intensity distribution along the Debye--Scherrer rings without requiring segmentation of individual diffraction peaks, texture tomography is well suited for fine-grained and highly deformed polycrystals. However, existing formulations have mainly been demonstrated on weakly textured materials with relatively smooth orientation distributions (Figure~\ref{regime_plot}C). For sharply textured materials, the support of the ODF is localized to a small set in orientation space and the diffraction signal transitions toward isolated peaks (Figure~\ref{regime_plot}B). Representing such an ODF accurately with a uniform discretization requires a large number of basis functions, since the number of orientation-space grid points scales cubically with angular resolution. Consequently, both the computational cost and memory requirements scale cubically with the angular resolution. In addition, the computational cost scales linearly with the number of voxels in the reconstructed volume. Together, these scaling properties limit the achievable angular and spatial resolution. Previous work reported an angular resolution of $3^\circ$ on a 150×150 spatial grid \citep{carlsen2025a}. However, for many materials science applications, the angular resolution needed is an order of magnitude higher \citep{Zelenika2024}. This leaves a portion of the spectrum uncovered: samples with sharp texture where current texture tomography implementations fail due to computational limits, yet where the diffraction signal is not sufficiently discrete for s3DXRD. Many practically important materials fall in this middle ground, including highly deformed metals with intragranular orientation gradients and geological samples with strong crystallographic preferred orientations. 

Bridging this gap is the central contribution of this work. We apply peak segmentation and indexing to identify candidate crystallographic orientations from the diffraction images \citep{hayashi2015, Hayashi2019, henningsson2024, wright2026}. This provides a natural way to restrict the reconstruction to the relevant regions of orientation space rather than discretizing it uniformly. We construct an adaptive basis for texture tomography by placing Gaussian basis functions only at orientations identified through indexing. This reduces the number of required basis functions, while retaining the high angular resolution, as the basis functions are centered around orientations known \emph{a priori} through peak indexing to be present in the sample.

Our second contribution is a GPU-based implementation of the forward model and optimization, designed for low memory consumption. This makes high-resolution reconstruction feasible on modern GPUs and scales naturally to full three-dimensional problems. The implementation is available as an open-source Python library \citep{caroe2026}.

Together, these contributions extend texture tomography into the discrete diffraction regime, bridging the gap between isotropic diffraction and discrete diffraction regimes, enabling spatially resolved ODF reconstruction in polycrystals with strongly localized texture. 

The remainder of the paper is structured as follows. In Section \ref{sec:model}, the texture tomography forward model is formulated; in Section \ref{sec:contributions}, the adaptive grid and computational outline are described; in Section \ref{sec:simulation}, the grid-adaptive Texture tomography is validated on a simulated dataset; in Section \ref{sec:experiment}, the method is applied to diffraction data taken on a tensile Al1050 specimen measured at the DanMAX beamline at MAX IV. Lastly, we discuss the findings and conclude in Sections \ref{sec:discussion} and \ref{sec:conclusion}.

\section{Texture tomography model}  \label{sec:model}

\begin{figure}[tbp] %
\begin{center}
\includegraphics[width=\textwidth]{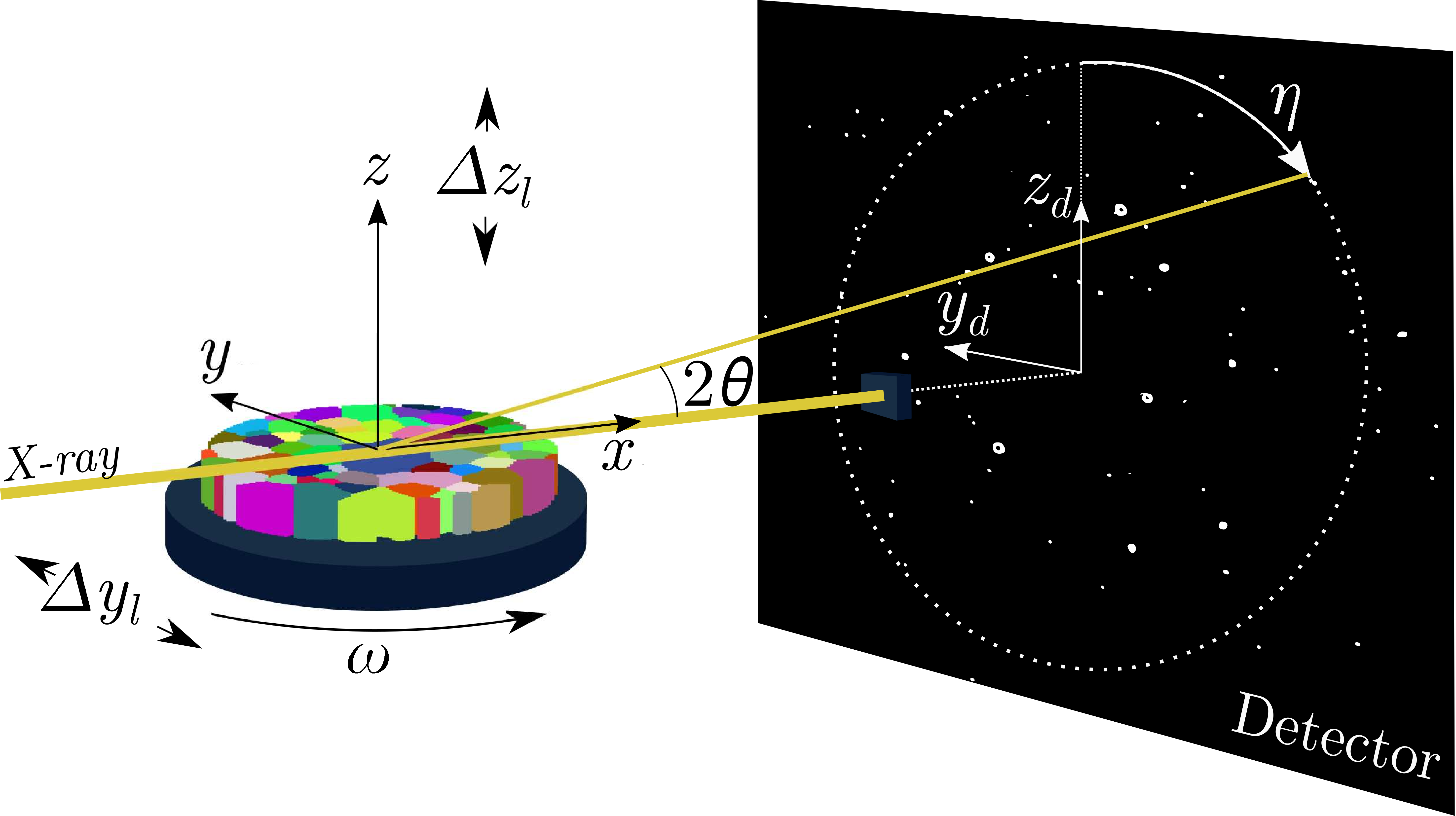} % NB use pdflatex for non-postscript
\end{center}
\caption{Experimental setup. The sample is translated in the $\mathbf{y}$--$\mathbf{z}$ plane with horizontal step size $\Delta y_l$ and vertical step size $\Delta z_l$. At each translation position, the sample is rotated through an angle $\omega$ about $\mathbf{z}_l$; the stacked detector frames illustrate that one image is collected per rotation step.} % NB \protect\citep{...} is required in floating figures
\label{setup}
\end{figure}

The experimental geometry used in this paper is that of single-axis texture tomography, which is identical to s3DXRD introduced by \citet{hayashi2015}. A polycrystalline sample is illuminated by a focused monochromatic X-ray pencil beam and scanned across the beam in the horizontal ($\mathbf{y}$) and vertical ($\mathbf{z}$) directions with step sizes $\Delta y_l$ and $\Delta z_l$, respectively, providing 2D spatial coverage of the cross-section. At each translation position, the sample is rotated through an angle $\Delta \omega$ about $\mathbf{z}$ while diffraction intensities are recorded on an area detector (Figure~\ref{setup}). The intensity at each detector pixel is described by the azimuthal angle $\eta$ and scattering angle $2\theta$; its variation along the Debye--Scherrer rings contains information about the orientation distribution of crystallites along the beam path.

To reconstruct the spatially varying texture from these measurements, we adopt the texture tomography framework of \citet{frewein2024} and \citet{carlsen2025a}. The texture tomography framework permits sample tilt, although all measurements in this work were acquired without tilt. The crystallographic texture at each spatial position is described by an ODF $\rho(\mathbf{U})$ where $\mathbf{U}\in\mathrm{SO}(3)$. Following \citet{carlsen2025a}, the ODF is represented as a non-negative expansion of symmetry-adapted Gaussian basis functions,
\begin{equation}
\rho(\mathbf{U}) = \sum_k c_k \rho_k(\mathbf{U}), \qquad
\rho_k(\mathbf{U}) =
C \sum_{\mathbf{G}\in\mathcal{G}}
\exp\left(
\frac{\mathrm{tr}(\mathbf{U}^\top \mathbf{G}\mathbf{U}_k)-1}
{2\sigma_k^2}
\right),
\end{equation}
where $\mathbf{U}_k$ denotes the center orientation of the $k$-th basis function and $\sigma_k$ controls its angular width. The set $\mathcal{G}$ contains the crystallographic point group symmetry operations, and the sum over $\mathbf{G}$ enforces symmetry equivalence of orientations. The constant $C$ is chosen such that each basis function is properly normalized over orientation space, and the coefficients $c_k \ge 0$ determine the local density of the ODF. Spatially varying texture is introduced into the model by partitioning real space into a voxel grid. The expansion coefficients $c_k$ then become voxel-dependent, $c_{kxyz}$. The operator that maps the ODF coefficients $c_{kxyz}$ to the diffraction intensity is linear.
Consequently, we stack all recorded diffraction measurements (i.e. azimuthally binned detector pixels over all rotations and scan positions) into a single data vector $\mathbf{I}$, and all expansion coefficients into a vector $\mathbf{c}$, so that the forward model takes the form
\begin{equation}
\label{eq:linear_system}
\mathbf{I} = \mathbf{A}\mathbf{c},
\end{equation}
where the forward operator decomposes as $\mathbf{A} = \mathbf{B}\mathbf{P}$, with $\mathbf{P}$ representing the tomographic X-ray projection in real space and $\mathbf{B}$ representing the pole figure transform that maps orientation basis functions to diffraction intensities through the crystal structure. The detailed construction of $\mathbf{A}$ is given in Appendix~\ref{appendix:forward_model}.

Given a diffraction dataset \(\mathbf{I}\), we solve the inverse problem of obtaining the ODF coefficients, \(\mathbf{c}\), by minimizing the Huber loss of the residual \(\mathbf{BP}\mathbf{c} - \mathbf{I}\) \citep{huber1964} subject to a non-negativity constraint,  \(\mathbf{c} \geq \mathbf{0}\), using the FISTA optimization algorithm \citep{beck2009}. No explicit regularization is applied; instead, early stopping acts as an implicit regularization strategy, exploiting semi-convergence in the iterative solution \citep{hansen2010}.

\section{Adaptive basis selection and computational pipeline} \label{sec:contributions}

\begin{figure}[tbp] %
\begin{center}
\includegraphics[width=0.5\textwidth]{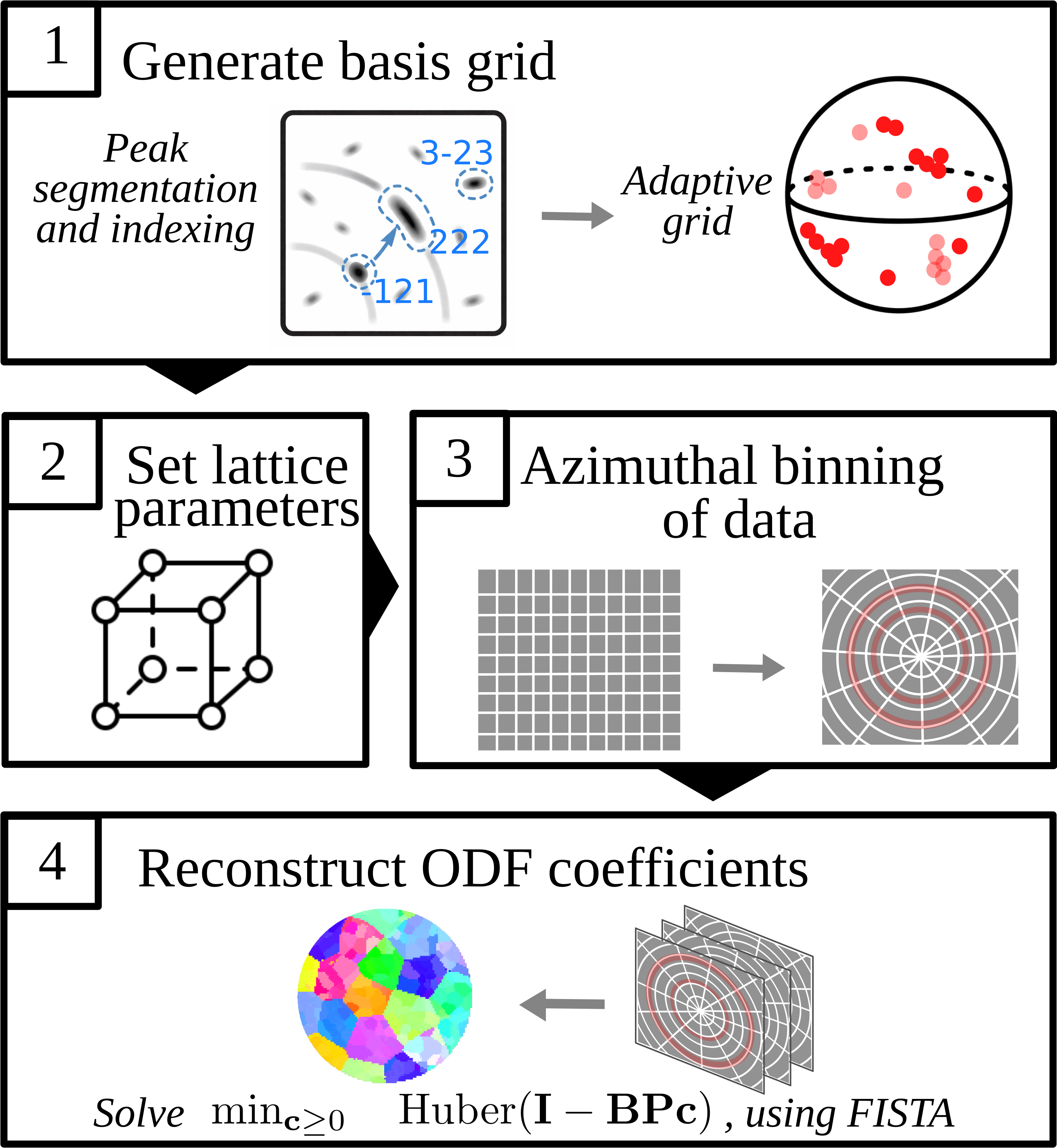} % NB use pdflatex for non-postscript
\end{center}
\caption{Overview of the reconstruction pipeline using an adaptive basis. (1) The orientation-space grid is constructed via peak segmentation and indexing. (2) Lattice parameters for the crystallographic phase are specified. (3) Raw detector images are azimuthally binned and normalized to produce the input data for the reconstruction. (4) The reconstruction is performed by minimizing the Huber loss of the residuals. The optimization yields ODF coefficients for each spatial voxel.}
\label{flow_chart}
\end{figure}

An overview of the reconstruction pipeline is shown in Figure~\ref{flow_chart}; each stage is described in detail in the following subsections.

\subsection{Data preprocessing}\label{sec:preprocessing}

Prior to reconstruction, diffraction rings are selected and azimuthally binned using PyFAI \citep{ashiotis2015}. Each ring is normalized by its total integrated intensity over all detector frames, removing the influence of structure factors.

\subsection{Basis grid selection}

The choice of basis function grid $\mathbf{U}_k$ and their widths $\sigma_k$ determines the angular resolution of the reconstruction. We consider two strategies.

\paragraph{Uniform grid.}
As a baseline, center orientations $\mathbf{U}_k$ are sampled at random from a uniform distribution in the fundamental zone of $\mathrm{SO}(3)$. The basis set is pruned to enforce a minimum angular separation between grid points. The widths $\sigma_k$ are chosen proportional to this spacing. This provides uniform coverage of orientation space but does not exploit any prior knowledge of the sample texture.

\paragraph{Adaptive grid.}
For sharply textured samples, the majority of the uniformly sampled basis functions contribute negligibly to the reconstruction. Rather than sampling orientation space uniformly, we place basis functions at orientations identified by the ImageD11 package \citep{wright2026} as present in the bulk sample. This is done by segmenting diffraction peaks on the raw detector images, extracting corresponding scattering vectors from two selected Debye-Scherrer rings to identify a candidate set of orientations, which is then refined against the full dataset. The resulting orientations serve as a basis function grid, with widths $\sigma_k$ matched to the observed peak widths. This yields a basis restricted to the relevant regions of orientation space, reducing the number of degrees of freedom while improving angular resolution.

\subsection{GPU implementation}

As introduced in Section~\ref{sec:model}, the forward operator can be written in separable form as $\mathbf{A} = \mathbf{B}\mathbf{P}$, where $\mathbf{P}$ is the tomographic projection---implemented using the Gratopy library \citep{bredies2021a, bredies2021b}---and $\mathbf{B}$ is a matrix with elements $B_k$, described in Appendix~\ref{appendix:forward_model}. For large basis sets, storing $\mathbf{B}$ explicitly is impractical; instead, the matrix is evaluated in batches during each forward and adjoint operation, keeping GPU memory usage independent of the total number of basis functions. With this batched implementation, the memory bottleneck is storing the basis function coefficients $c_{kxyz}$ rather than storing the much larger matrix $\mathbf{B}$. The full pipeline is implemented in Python and OpenCL. The code is available as an open-source library on GitHub~\citep{caroe2026}.

\section{Simulated dataset}\label{sec:simulation}
In order to demonstrate that our method can operate in both the discrete diffraction and strong texture regimes, we perform two numerical simulations featuring polycrystals with a high and low degree of intragranular orientation spread respectively. To benchmark the accuracy of the reconstructions, we compare the texture tomography (TT) reconstructions with uniform and adaptive basis function grids (TT-Uniform and TT-Adaptive, respectively) against the s3DXRD point-by-point method (PBP) described in \citet{henningsson2024}.

\subsection{Setup and reconstruction}
We validate the method displayed in Figure \ref{flow_chart} on a simulated aluminum (fcc) polycrystal generated with \texttt{xrd\_simulator} \citep{henningsson2023}. The simulated polycrystal consists of 20~grains with mean orientations drawn from a uniform random distribution on $\mathrm{SO}(3)$, each subdivided into orientation domains whose orientations deviate linearly from the grain mean along a randomly chosen spatial direction, up to a maximum of $1^\circ$ or $10^\circ$ mosaicity, representing a nearly perfect and a moderately deformed polycrystal, respectively. A spatially varying elastic strain field is additionally imposed to test robustness to strain-induced peak shifts present in real experiments; it is constructed as a self-equilibrated Gaussian-process stress field with amplitudes of approximately $20\,\mathrm{MPa}$ and a spatial correlation length of $1.5\,\upmu\mathrm{m}$. Full simulation details and ground-truth field visualizations are given in Appendix~\ref{appendix:simulation}. Diffraction is simulated in the ideal-crystal limit, placing peaks at the exact Laue condition positions, accounting for structure factors, the Lorentz factor, and horizontal polarization; absorption is neglected.

A single slice is scanned during rotation about a single vertical axis, using $N_y = 103$ translations, $N_\omega = 360$ rotations in steps of $\Delta\omega = 0.5^\circ$, and a beam energy of $50\,\mathrm{keV}$. The data are integrated into $N_\eta = 180$ azimuthal bins and 8 radial regions centered at the scattering angles of the first 8 non-extinct Miller index families and normalized per ring as described in Section~\ref{sec:preprocessing}. For both the $1^\circ$ and $10^\circ$ mosaicity samples, the uniform grid uses $\sigma_k = 1.3^\circ$ and the adaptive grid uses $\sigma_k = 0.4^\circ$, where $\sigma_k$ denotes the angular width parameter of the Gaussian basis functions on $\mathrm{SO}(3)$. The number of basis functions for each case is indicated in Figures~\ref{simulation_figure_10deg} and~\ref{simulation_figure_1deg}.

To display the reconstructed ODF in each voxel, we extract a representative orientation defined as the orientation that minimizes the coefficient weighted sum of misorientation angles to all ODF components, and display the result as an inverse pole figure (IPF) map. The IPF maps are computed using the Orix library \citep{anes_2026, johnstone2020}. The misorientation between the ground truth and representative orientation is also calculated. Additionally, we extracted grain and subgrain boundaries following the comparison metric introduced in \citet{kim2023}, which identifies boundaries from kernel average misorientation (KAM) maps, and we compare the reconstructed grain boundaries to the ground truth. The grain and subgrain boundaries were extracted using a $4^\circ$ and $0.5^\circ$ threshold on the KAM respectively.

\subsection{Results}

\begin{figure}[tbp] %
\begin{center}
\includegraphics[width=0.9\textwidth]{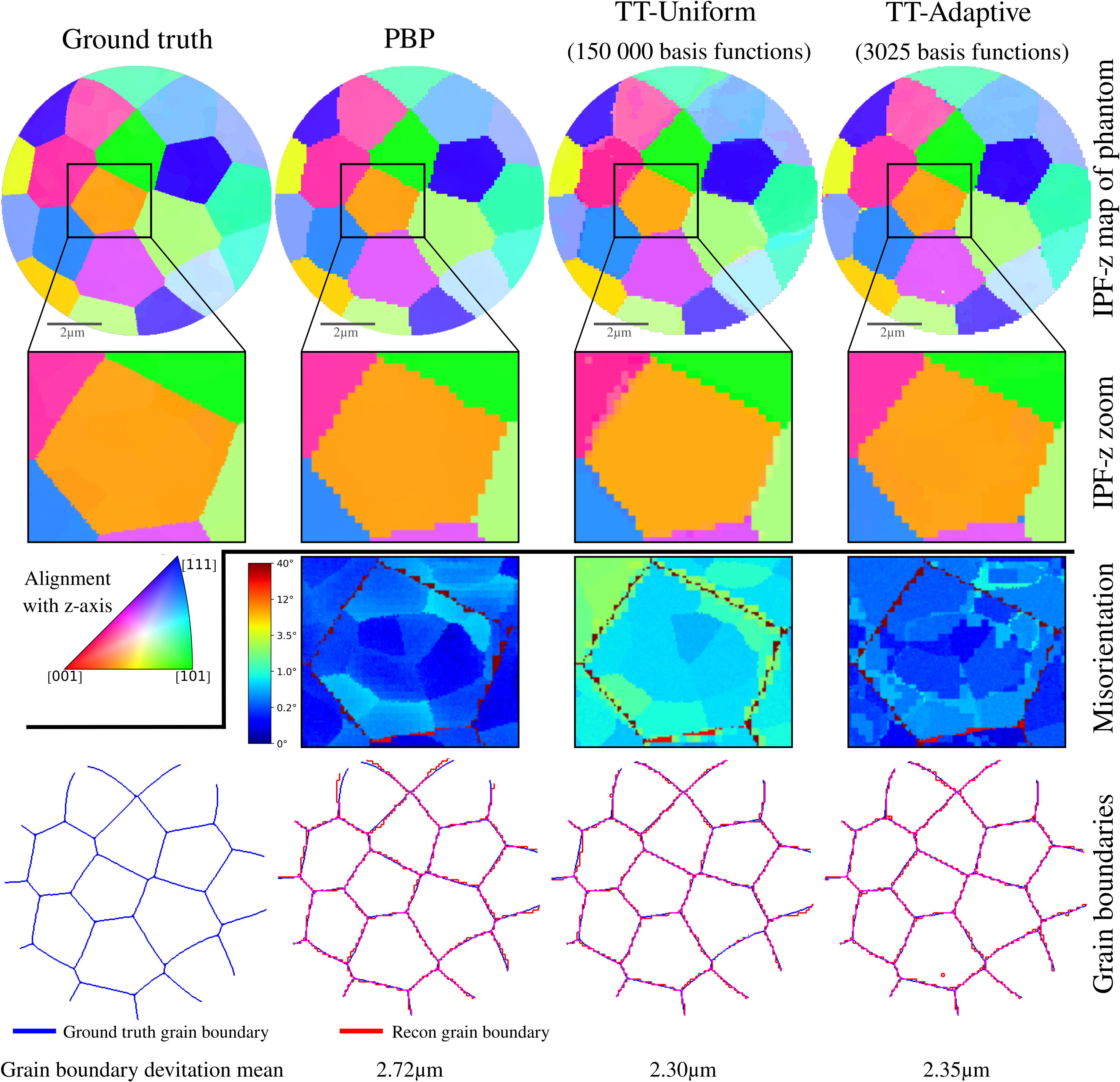} % NB use pdflatex for non-postscript
\end{center}
\caption{Simulation results for the $1^\circ$ mosaicity sample.  The ground truth shown in the left column is compared with the baseline point-by-point algorithm in the second column, and the two texture tomography algorithms in the third and fourth columns. First and second row: IPF-Z maps. Third row: Misorientation between ground truth and reconstruction. Fourth row: Reconstructed and ground truth grain boundaries.}
\label{simulation_figure_1deg}
\end{figure}

\paragraph{$1^\circ$ mosaicity.}
Figure~\ref{simulation_figure_1deg} shows the IPF-Z and misorientation maps for the $1^\circ$ mosaicity sample. As expected for a nearly perfect-crystal sample with minimal peak overlap, the PBP method performs well. Both TT reconstructions also recover the grain structure accurately, although TT-Uniform exhibits larger intragranular errors. Peak segmentation and indexing required 21 minutes. This step provides the input for both PBP and TT-Adaptive reconstructions, while TT-Uniform operates directly on the raw diffraction data and does not rely on this intermediate representation. The PBP reconstruction incurs no additional computational cost beyond segmentation and indexing. TT-Adaptive uses only $3025$ basis functions, compared with $150000$ for TT-Uniform, reducing the reconstruction time from 225 minutes to 5 minutes.

\begin{figure}[tbp] %
\begin{center}
\includegraphics[width=0.9\textwidth]{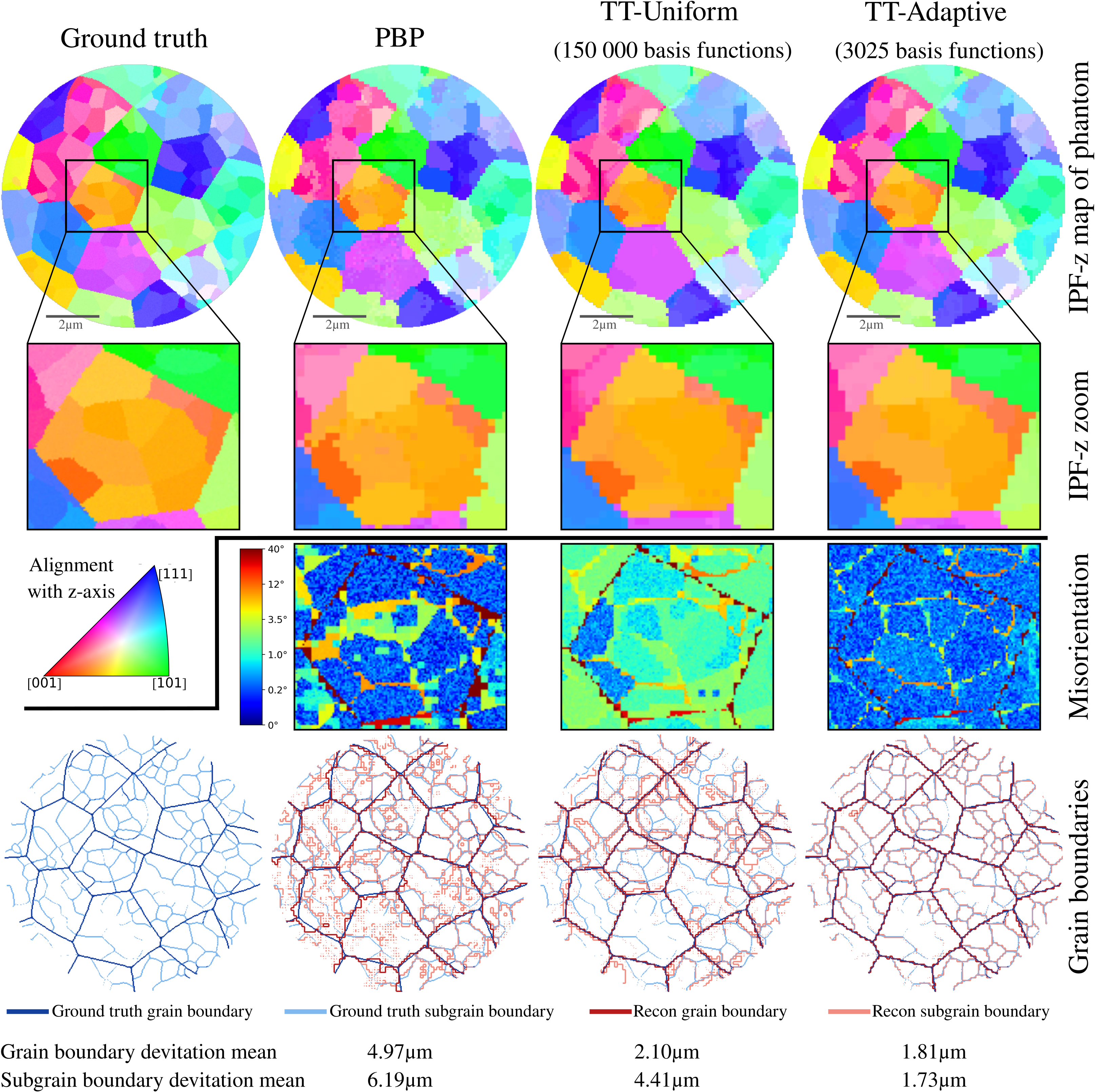} % NB use pdflatex for non-postscript
\end{center}
\caption{Simulation results for the $10^\circ$ mosaicity case. The ground truth shown in the left column is compared with the baseline point-by-point algorithm in the second column, and the two texture tomography algorithms in the third and fourth columns. First and second row: IPF-Z maps. Third row: Misorientation between ground truth and reconstruction. Fourth row: Reconstructed and ground truth grain boundaries.}
\label{simulation_figure_10deg}
\end{figure}

\paragraph{$10^\circ$ mosaicity.}
Figure~\ref{simulation_figure_10deg} shows the IPF-Z and misorientation maps for the $10^\circ$ mosaicity sample. The PBP baseline recovers the overall grain structure but produces irregular grain boundaries, with boundary voxels exhibiting large misorientation errors. TT-Uniform reconstructs straight grain boundaries, but the intragranular misorientation is limited by the spacing of the uniform grid on $\mathrm{SO}(3)$, since the reconstruction accuracy cannot substantially exceed the angular distance between neighboring basis functions. TT-Adaptive resolves both the grain boundaries and the intragranular orientation gradients, yielding lower misorientation throughout the grain interior and even resolving subgrain boundaries. The reconstruction times are similar to those of the $1^\circ$ sample: 225 minutes for TT-Uniform and 5 minutes for TT-Adaptive. In this case, peak segmentation and indexing required 38 minutes, while all other computational costs remained unchanged compared to the $1^\circ$ case.

\section{Experimental demonstration on a deformed tensile Al specimen} \label{sec:experiment}

We apply basis-adaptive texture tomography to an experimental dataset of a tensile-deformed aluminum specimen. The sample is 1050 aluminum alloy ($99.5\%$ pure), cold rolled and subsequently heat treated to induce recrystallization. It was deformed to a $15\%$ tensile elongation using the load frame D-stroi \citep{frankus2026}. The engineering stress--strain curve can be found in Appendix~\ref{appendix:sample_deformation}. The rolling direction (RD) is aligned with the $Z$ sample axis (the tensile direction), while the normal direction (ND) and transverse direction (TD) lie in the $XY$ plane. Prior to tensile deformation the sample had a $1.2\rm{mm}$ by $1.2\rm{mm}$ cross section and a length of $10\rm{mm}$. The average grain size (mean chord length) is $70\upmu \mathrm{m}$ \citep{knudsen2006}. This specimen provides a relevant test case for basis-adaptive texture tomography, since the plastic deformation introduces intragranular orientation spread and partially overlapping diffraction peaks while having a comparatively simple single-phase microstructure with well-understood deformation behavior.

\subsection{Experimental setup}

\begin{figure}[tbp] %
\begin{center}
\includegraphics[width=\textwidth]{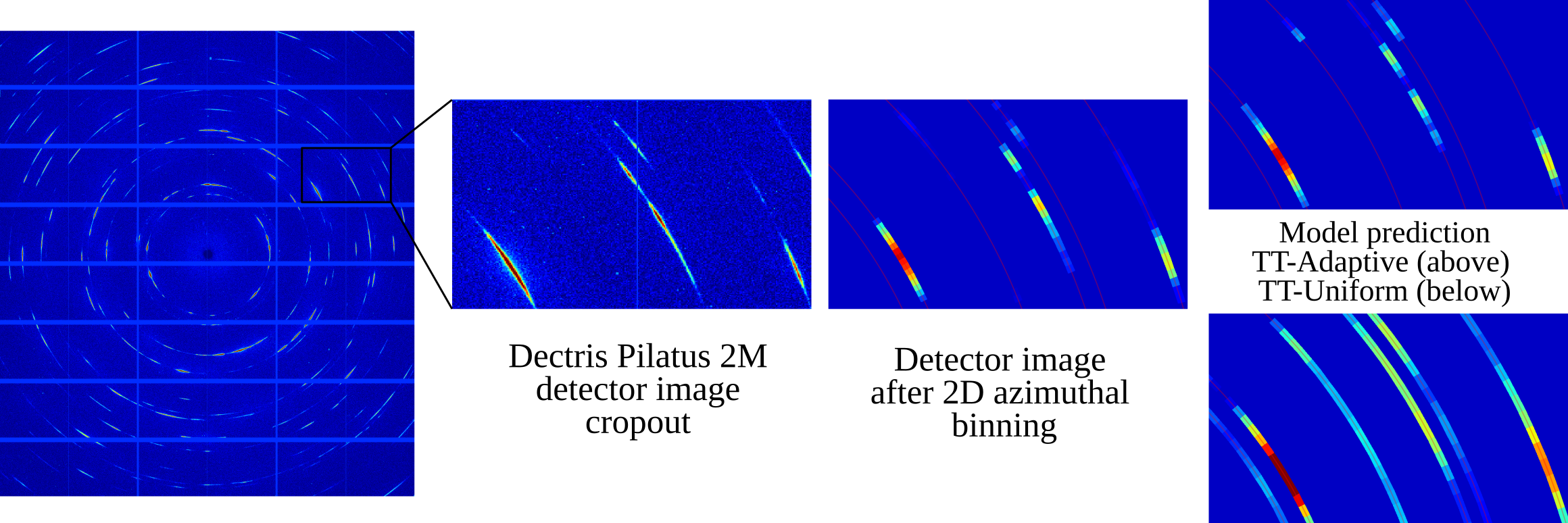} % NB use pdflatex for non-postscript
\end{center}
\caption{Comparison between data and forward model predictions. From left to right: a full detector image from the aluminum specimen experiment; a zoomed view of a selected region; the same region after azimuthal integration onto a polar grid; and the corresponding model predictions using the adaptive grid ($37\,599$ basis functions, $\sigma_k = 0.4^\circ$) and a uniform grid ($150\,000$ basis functions, $\sigma_k = 1.3^\circ$).}
\label{detector_figure}
\end{figure}

A single slice was scanned at the DanMAX beamline at MAX IV using $35\,\mathrm{keV}$ X-rays and a DECTRIS PILATUS3 2M detector placed $24.49\,\mathrm{cm}$ from the sample. The beam was focused to a $10\,\upmu\mathrm{m}$ spot size, which is also the translation step size. The sample was rotated continuously through a full $360^\circ$ including both $\omega=0^\circ$ and $\omega=360^\circ$, acquiring 3003 detector frames per translation position. A representative detector image and the corresponding detector image predicted from the reconstruction by the forward model are shown in Figure~\ref{detector_figure}.

\subsection{Reconstruction}

The diffraction data was integrated into $N_\eta = 540$ azimuthal bins across the 9~lowest-angle non-extinct reflection classes, and normalized per ring as described in Section~\ref{sec:preprocessing}. The rotation data was binned to $N_\omega=500$ steps to reduce computational cost, with no significant change observed at higher angular sampling. For the reconstruction, we used $37\,599$ basis functions obtained from peak indexing with a width of $\sigma = 0.4^\circ$. The code used in the reconstruction is publicly available \citep{caroe2026b}.

As in Section~\ref{sec:simulation}, a representative orientation is extracted in each voxel and displayed as an IPF map. The Kernel Average Misorientation (KAM) is also computed, providing a voxelwise measure of local orientation spread that highlights grain boundaries. IPF and KAM maps are computed using the Orix library \citep{anes_2026, johnstone2020}.

\subsection{Results}

\begin{figure}[tbp] %
\begin{center}
\includegraphics[width=\textwidth]{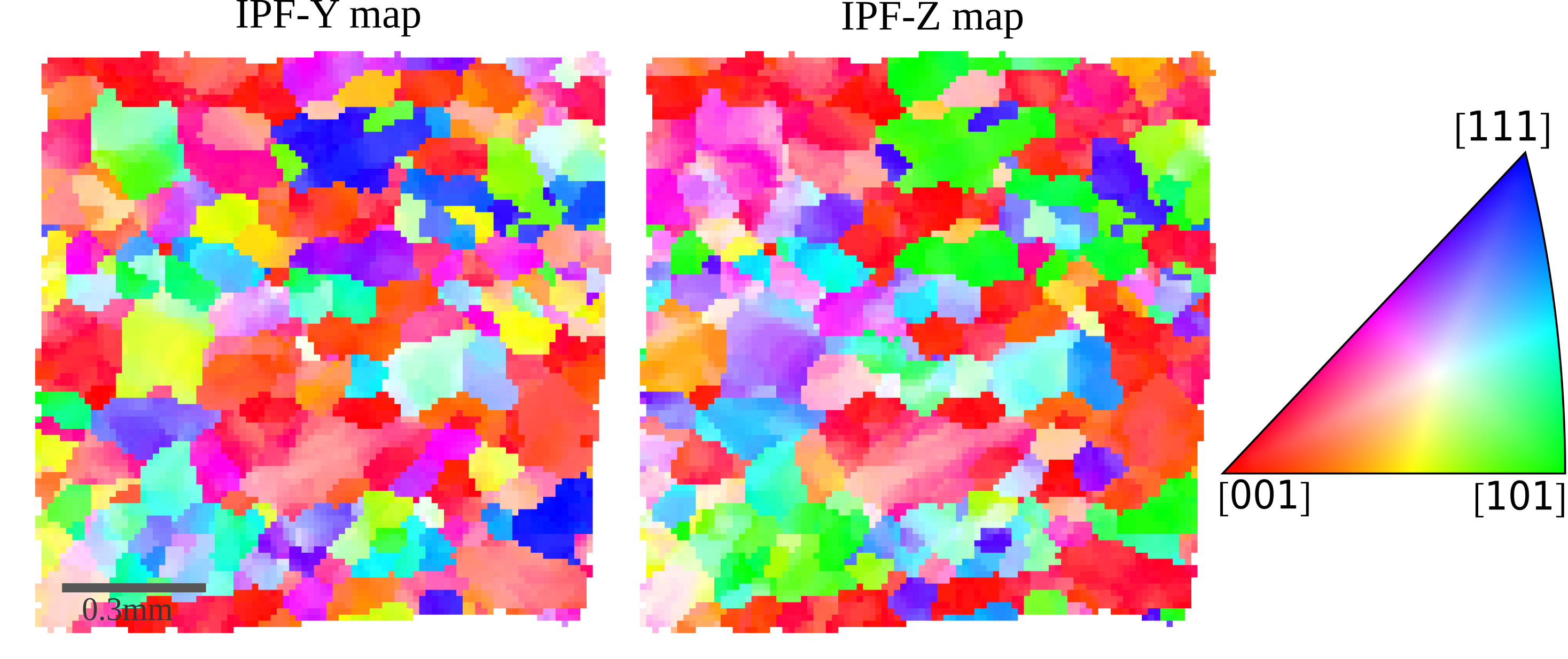} % NB use pdflatex for non-postscript
\end{center}
\caption{Inverse pole figure (IPF) maps of the reconstructed Al1050 cross-section. Each pixel displays the representative orientation colour-coded by the standard cubic IPF colour key, projected along the $Y$ (left) and $Z$ (right) sample axes.}
\label{ipf_maps}
\end{figure}

Figure~\ref{ipf_maps} shows IPF maps of the reconstructed cross-section along the $Y$ and $Z$ sample axes. Approximately 100~grains are resolved, consistent with the $70\,\upmu\mathrm{m}$ mean chord length across the $1.2\,\mathrm{mm}$ cross-section. Within individual grains, intragranular orientation gradients of a few degrees are visible, in line with the moderate deformation of the sample.

\begin{figure}[tbp] %
\begin{center}
\includegraphics[width=\textwidth]{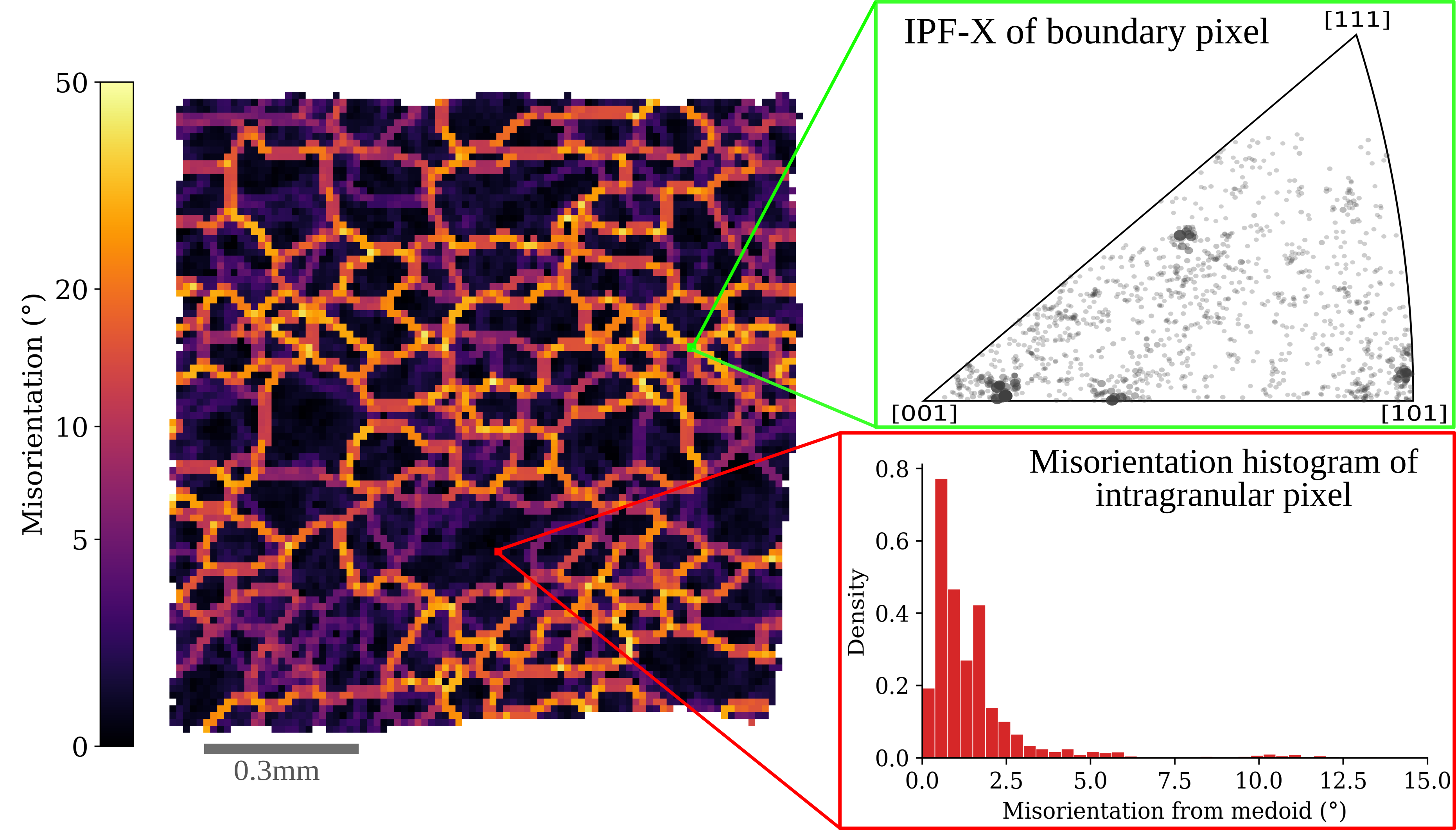} % NB use pdflatex for non-postscript
\end{center}
\caption{Left: Kernel Average Misorientation (KAM) map of the same cross-section as in Figure~\ref{ipf_maps}, highlighting grain and subgrain boundaries. Top right: inverse pole figure of the reconstructed ODF at a selected grain boundary pixel; each cluster corresponds to a neighboring grain. Bottom right: histogram of misorientation from the representative orientation for a selected intragranular pixel.}
\label{misorientation_figure}
\end{figure}

Figure~\ref{misorientation_figure} shows a KAM map together with ODF visualizations for two selected voxels. The KAM map reveals both high-angle grain boundaries and lower-angle subgrain boundaries, with orientation differences of a few degrees across the subgrain interfaces, characteristic of the dislocation substructure that develops in fcc metals at this strain level \citep{humphreys2004}. At a selected grain boundary voxel, the reconstructed ODF displays four distinct orientation components---one per neighboring grain. Because the method reconstructs a full ODF rather than forcing a single orientation per voxel, it naturally captures this multi-grain contribution. At a selected intragranular voxel, the ODF density is concentrated within $\sim\!2.5^\circ$ of the representative orientation, consistent with the presence of deformation-induced lattice rotations within the strained specimen.

The reconstruction further indicates that individual grains are partitioned into subgrains with orientation spreads of a few degrees. Such orientation variations are expected to produce closely spaced sub-peaks in the diffraction images corresponding to slightly misoriented subgrains.

\begin{figure}[tbp] %
\begin{center}
\includegraphics[width=0.99\textwidth]{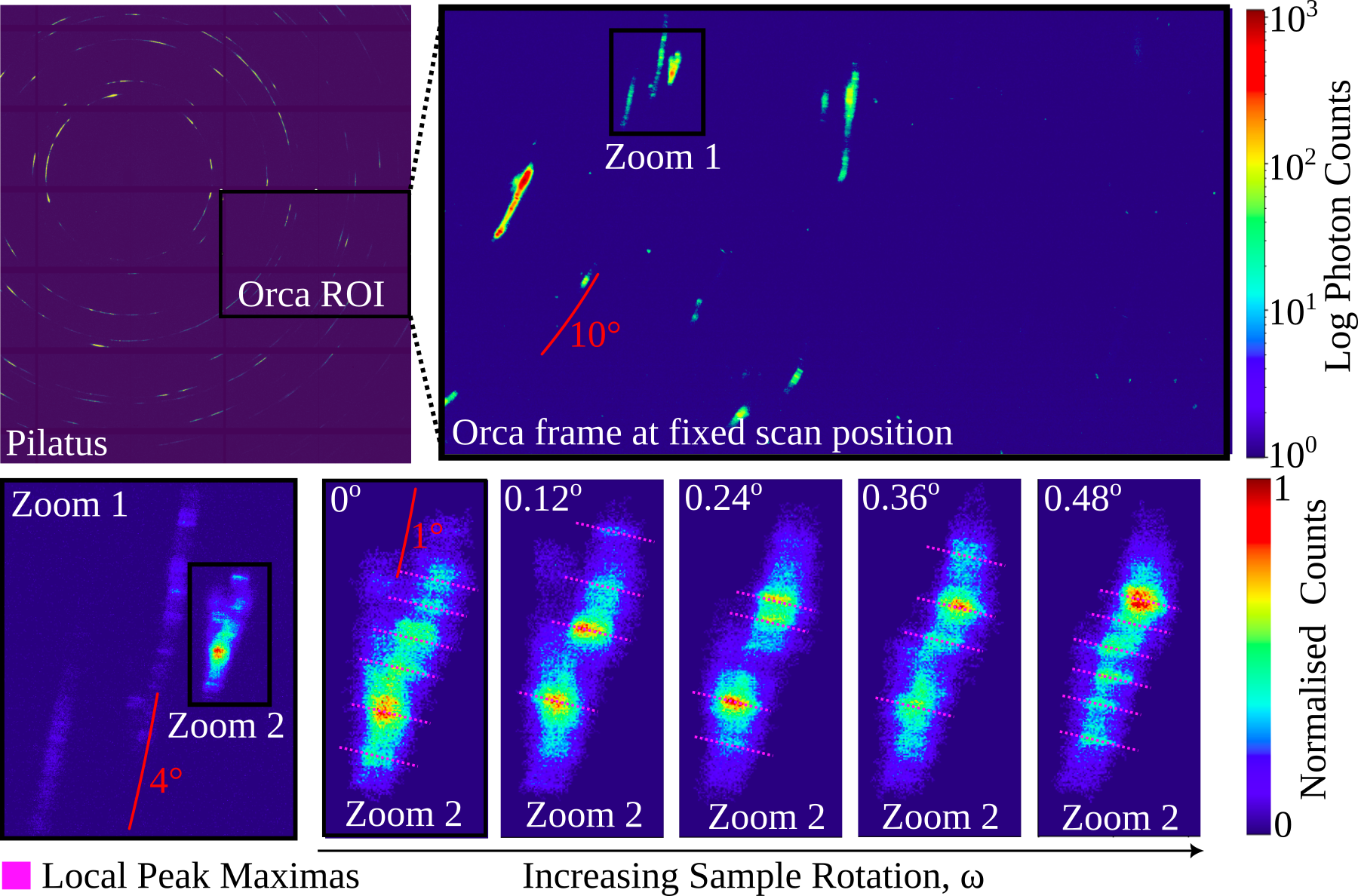} % NB use pdflatex for non-postscript
\end{center}
\caption{Comparison of PILATUS and ORCA detector images. (A):~PILATUS detector image at the central scan position with the ORCA field of view (ROI) annotated. (B):~Corresponding ORCA image (pixel size $1.1\,\mu\mathrm{m}$, detector distance $12\,\mathrm{mm}$), compared to the PILATUS pixel size of $172\,\mu\mathrm{m}$ at $244.9\,\mathrm{mm}$. (C):~ Zoomed in view on a particular region of the ORCA image. (D):~Further zoomed in view on a particular peak tracked as a function of rotation angle $\omega$, revealing several distinct local maxima (annotated by pink dashed lines), indicating multiple orientation components within the illuminated volume.}
\label{orca}
\end{figure}

To investigate this experimentally, complementary high-resolution diffraction images were collected at DanMAX using a Hamamatsu ORCA camera with approximately 20 times higher detector resolution than the PILATUS detector, however with a smaller field of view. The ORCA images were acquired after the PILATUS scan without dismounting the sample, allowing direct comparison between the detector images, as shown in Figure~\ref{orca}. Owing to the higher detector resolution, reflections that appear as single merged peaks in the PILATUS images are resolved into multiple local maxima in the ORCA data as a function of the sample rotation angle, $\omega$. Although the basis-adaptive texture tomography reconstruction was performed using only the lower-resolution PILATUS data, the experimentally observed peak splitting is consistent with the reconstructed subgrain structure and intragranular orientation spread.

\section{Discussion and outlook} \label{sec:discussion}
The results (Figures \ref{simulation_figure_1deg}, \ref{simulation_figure_10deg}, \ref{ipf_maps} and \ref{misorientation_figure}) show that the adaptive basis formulation is capable of resolving intragranular orientation structures at high angular resolution. By restricting the reconstruction to orientations extracted from peak segmentation and indexing, the number of basis functions can be reduced substantially compared with a fine uniform discretization of orientation space. This reduction in computational complexity makes it possible to increase either the spatial resolution or reconstruct larger samples while retaining high angular resolution.

The reconstruction of the aluminum specimen dataset completed in 68~minutes using basis-adaptive texture tomography with an adaptive grid consisting of 37599 grid points. The adaptive grid was obtained by pruning the list of orientations returned by the peak segmentation and indexing steps, enforcing a minimal distance between grid points of $1\,\mathrm{mrad}$. However, to get an overview of the sample texture during beamtime, a high-resolution reconstruction is likely not necessary. By enforcing a larger minimal distance between grid points and increasing the basis function width $\sigma$, the basis set can be considerably reduced. As the reconstruction time scales linearly with the number of basis functions, coarse reconstructions can be completed within a few minutes during beamtime, with reconstruction quality depending on the chosen basis grid.

The adaptive basis used in this article was constructed from peak segmentation and indexing, but other strategies for populating orientation space are possible. For example, by downsampling the spatial voxel grid to a few or even to a single voxel, computational resources are freed up and can be used to create a very fine uniform grid in orientation space. The texture tomography algorithm then returns a bulk sample ODF describing the overall texture of the sample. One can then draw samples at random from this ODF, creating a set of basis function grid that is denser in regions where the ODF has higher density. This basis function grid can then be used in the highly spatially resolved reconstruction.

Another strategy for choosing the basis set would be by using iterative refinement similar to the strategy used in \citet{henningsson2023b}. This could be done by making an initial coarse reconstruction with large basis width, $\sigma_k$. The basis set would then be expanded by discarding regions of orientation space where the reconstructed ODF has diminishing density. In regions close to non-discarded grid points, a finer grid would be sampled, and a reconstruction made using the new grid with smaller $\sigma_k$. By running a few iterations, one would obtain a basis function grid adapted to the orientations that are present in the sample.

Apart from an adaptive basis function grid, one could also imagine optimizing the basis function widths $\sigma_k$. In this article, we have only considered a $\sigma_k$ that is constant for all basis functions. However, allowing for a variable $\sigma_k$, we would be able to represent the whole spectrum of textures from uniformly random to sharp. This is especially relevant for geological samples, where both powderous regions and large grains can be simultaneously present.

Relevant for geological materials is also the presence of multiple phases within the sample. The current basis-adaptive texture tomography framework described in this article only allows for specifying one set of lattice parameters; however, in future work, we will extend the framework to multiphase materials by constructing independent basis sets for each crystallographic phase, each with its own lattice parameters and structure factors. This extension is trivial if the phases have non-overlapping diffraction rings; however, in case of overlap, the model needs to simultaneously reconstruct ODFs for multiple phases.

\section{Conclusion} \label{sec:conclusion}
Polycrystalline materials such as plastically deformed metals, ferroelastic materials, and geological aggregates with strong preferred orientations can produce spotty, overlapping diffraction patterns in s3DXRD-type experiments that are not well described by existing reconstruction algorithms. In this regime, point-by-point methods become unreliable due to the presence of too many diffraction peaks, while existing texture tomography algorithms require prohibitively fine orientation discretizations to represent sharply localized orientation distributions.

We propose basis-adaptive texture tomography to address this gap by replacing the uniform orientation discretization of texture tomography with an adaptive basis constructed from orientations identified directly from the diffraction data. This combines the angular selectivity of peak-based indexing with the ability of texture tomography to reconstruct a full orientation distribution function in every voxel, allowing multiple orientations, subgrains, or domains to coexist locally rather than enforcing a single grain assignment. By concentrating the degrees of freedom in the regions of orientation space occupied by the sample texture, sharply localized orientation distributions can be reconstructed at tractable computational cost.

A GPU-accelerated forward model makes these reconstructions feasible within beamtime. Simulations and experimental measurements on tensile-deformed aluminum show improved reconstruction of grain and subgrain structures with intragranular orientation spreads of several degrees, opening texture tomography to a wider class of heterogeneous polycrystalline materials.

\appendix
\section{Forward model derivation}\label{appendix:forward_model}

\subsection{Experimental geometry}

The laboratory coordinate system is denoted by $(\mathbf{x}_l,\mathbf{y}_l,\mathbf{z}_l)$, with the X-ray beam parallel to $\mathbf{x}_l$ (Figure~\ref{setup}). The sample coordinate system is $(\mathbf{x}_\omega,\mathbf{y}_\omega,\mathbf{z}_\omega)$. A detector is positioned approximately perpendicular to the beam at a distance $D$ from the sample, and its position and orientation are determined through calibration.

During acquisition, the sample is translated in the $\mathbf{y}_l$–$\mathbf{z}_l$ plane and, for each translation, rotated about $\mathbf{z}_l$ with diffraction intensities recorded at discrete angular steps. The rotation of a point $\mathbf{v}$ in the sample frame to the laboratory frame is given by $\mathbf{v}_l = \mathbf{\Omega}\mathbf{v}_\omega$, where
\begin{equation}
\mathbf{\Omega} =
\begin{bmatrix}
\cos\omega & -\sin\omega & 0 \\
\sin\omega & \cos\omega  & 0 \\
0          & 0           & 1
\end{bmatrix}.
\end{equation}

Detector coordinates $(y_d,z_d)$ are expressed in polar form $(\eta,2\theta)$, where $\eta$ is the azimuthal angle and $2\theta$ the scattering angle. The corresponding scattering vector in laboratory coordinates is
\begin{equation}
\mathbf{G}_l = \frac{2\pi}{\lambda} \begin{bmatrix} \cos(2\theta)-1 \\ -\sin(2\theta)\sin\eta \\ \sin(2\theta)\cos\eta \end{bmatrix} = \frac{4\pi}{\lambda}\sin\theta \begin{bmatrix} -\sin\theta \\ -\cos\theta\sin\eta \\ \cos\theta\cos\eta \end{bmatrix}.
\end{equation}

\subsection{Diffraction intensity model}

For a polycrystalline sample with bulk texture described by an ODF $\rho(\mathbf{U})$ on $\mathrm{SO}(3)$, the diffraction intensity at rotation $\mathbf{\Omega}$ can be expressed as
\begin{equation}
 B_{\rho,\mathbf{\Omega}}( \mathbf{G}_l) = \sum_{\mathbf{Q}\in \bm{\Lambda}\mathbb{Z}^3} f_{\mathbf{Q}}\,\mathcal{P}_{\rho,\mathbf{Q}}(\mathbf{\Omega}^\top \mathbf{G}_l),
\end{equation}
where $\bm{\Lambda}$ is the reciprocal lattice matrix, $f_{\mathbf{Q}}$ are structure factors and $\mathcal{P}_{\rho,\mathbf{Q}}$ denotes the pole figure transform \citep{matthies1987}.

Expanding the ODF in a basis expansion, $\rho(\mathbf{U}) = \sum_k c_k \rho_k(\mathbf{U})$, the intensity becomes linear in the coefficients,
\begin{equation}
B_{\rho,\mathbf{\Omega}}( \mathbf{G}_l) = \sum_k c_k\, B_{\rho_k,\mathbf{\Omega}}( \mathbf{G}_l).
\end{equation}

For the Gaussian basis functions defined in the main text, the pole figure transform admits the closed-form approximation
\begin{equation}
\mathcal{P}_{\rho_k,\mathbf{Q}}(\mathbf{G}_l) \approx \frac{2}{\sigma_k^2} \exp\left(-\frac{1-\alpha(\mathbf{G}_l)}{2\sigma_k^2}\right),
\end{equation}
where $\alpha(\mathbf{G}_l) = \frac{\mathbf{Q}^T\mathbf{U}_k\mathbf{G}_l}{\|\mathbf{Q}\| \cdot \|\mathbf{G}_l\|}$. This expression is obtained by truncating the asymptotic expansion of a modified Bessel function that arises in the exact pole figure transform of the Gaussian kernel; see \citet{matthies1987} for the full derivation. The approximation is accurate when $\sigma_k$ is small relative to the angular separation between neighboring pole figure contributions. Note that the Lorentz factor is included in the model, whereas the polarization factor is not.

\subsection{Spatially resolved model}

For spatially varying texture with voxel-dependent coefficients $c_{kxyz}$, the measured intensity at rotation $\mathbf{\Omega}$ and translation $(t_y,t_z)$ is
\begin{equation}
I_{\mathbf{\Omega},t_y,t_z}(\mathbf{G}_l) = \sum_{xyz}\sum_k c_{kxyz}\,P_{xyz}^{t_y,t_z}\,B_k(\mathbf{\Omega}^\top \mathbf{G}_l),
\end{equation}
where $P_{xyz}^{t_y,t_z}$ represents the X-ray transform contribution of each voxel to the measurement at translation $(t_y,t_z)$. Collecting all terms yields the linear system $\mathbf{I} = \mathbf{A}\mathbf{c}$ used in the main text.

\section{Simulation Setup} \label{appendix:simulation}
Diffraction data were simulated from synthetic aluminum samples using
\texttt{xrd\_simulator}~\citep{henningsson2023}. Two samples were considered: a weakly deformed
sample with a small orientation spread, and a highly deformed sample with a
large orientation spread. Both samples had the same geometry, grain morphology,
elastic strain-generation procedure, and diffraction setup, but differed in the
prescribed domain-scale mosaicity, which were set to \(1^\circ\) and \(10^\circ\),
respectively. The aluminum phase was defined using a cubic unit cell with
\(a = 4.0493\,\mathring{A}\) and space group \(Fm\bar{3}m\). The sample geometry
was a thin cylinder with radius \(5\,\upmu\mathrm{m}\) and axial half-thickness
\(0.1\,\upmu\mathrm{m}\), discretized into 163677 tetrahedral elements.

The grain structure was generated from 20 randomly placed seed points in the
cylindrical cross-section. Rather than using a strict Euclidean Voronoi
tessellation, the grain labels were assigned using an RBF-interpolated
seed-label field, producing a curved Voronoi-like tessellation of the mesh.
Each grain was further subdivided into orientation domains using the same
RBF-based labelling procedure, with approximately 200 domains in total, and with
the number of domains in each grain proportional to its volume. The mean
orientation of each grain, \(R_g\), was drawn independently from the uniform
distribution on \(\mathrm{SO}(3)\). For each grain, a random unit vector
\(\mathbf{a}_g\) was drawn to define a spatial orientation-gradient direction.
If \(\mathbf{c}_{gd}\) is the centre of domain \(d\) in grain \(g\), and
\(\mathbf{c}_g\) is the grain centre, the normalized domain-centre coordinate was
defined as
\begin{equation}
    \hat{\mathbf{r}}_{gd}
    =
    \frac{\mathbf{c}_{gd}-\mathbf{c}_g}
         {\max_d \lVert \mathbf{c}_{gd}-\mathbf{c}_g\rVert}.
\end{equation}
The domain misorientation angle was then set to
\begin{equation}
    \theta_{gd}
    =
    m\, \mathbf{a}_g \cdot \hat{\mathbf{r}}_{gd},
\end{equation}
where \(m\) is the prescribed mosaicity, either \(1^\circ\) or \(10^\circ\). Each domain was assigned a random misorientation axis \(\mathbf{n}_{gd}\), and the domain-average orientation was then generated as
\begin{equation}
    R_{gd}
    =
    R(\mathbf{n}_{gd}, \theta_{gd}) R_g ,
\end{equation}
where \(R(\mathbf{n}_{gd}, \theta_{gd})\) denotes the rotation matrix
corresponding to a rotation of angle \(\theta_{gd}\) about the randomly chosen unit axis \(\mathbf{n}_{gd}\).

Finally, individual tetrahedra were assigned small random perturbations around
their domain-average orientation. For each tetrahedron, the perturbation axis
was drawn isotropically and the perturbation angle was drawn uniformly from
\([-m/20,m/20]\).

To generate mechanically realistic elastic strain tensor fields, an equilibrium-constrained Gaussian process (GP) stress field was used, similar to
the GP formulation of~\citep{Henningsson2021}. The construction was based on a Beltrami stress potential. The independent components of this potential were assigned GP priors using the same squared-exponential scalar kernel,
\begin{equation}
    k(\mathbf{x},\mathbf{x}')
    =
    \sigma_{\mathrm{GP}}^2
    \exp\!\left(
        -\frac{\lVert \mathbf{x}-\mathbf{x}' \rVert^2}{2l^2}
    \right),
\end{equation}
where \(l\) defines the spatial correlation length and \(\mathbf{x}\) is a spatial coordinate in the sample reference frame. The covariance of the six
Cauchy stress components was then obtained by applying the Beltrami differential operators to this kernel (see \citet{Henningsson2021} for details on this operator). The resulting random stress tensor field, \(\boldsymbol{\sigma}(\mathbf{x})\), is therefore self-equilibrated, that is, it satisfies
\begin{equation}
\begin{aligned}
    \frac{\partial \sigma_{xx}}{\partial x}
    + \frac{\partial \sigma_{xy}}{\partial y}
    + \frac{\partial \sigma_{xz}}{\partial z}
    &= 0, \\
    \frac{\partial \sigma_{yx}}{\partial x}
    + \frac{\partial \sigma_{yy}}{\partial y}
    + \frac{\partial \sigma_{yz}}{\partial z}
    &= 0, \\
    \frac{\partial \sigma_{zx}}{\partial x}
    + \frac{\partial \sigma_{zy}}{\partial y}
    + \frac{\partial \sigma_{zz}}{\partial z}
    &= 0,
\end{aligned}
\end{equation}
where
\begin{equation}
\boldsymbol{\sigma}
=
\begin{bmatrix}
\sigma_{xx} & \sigma_{xy} & \sigma_{xz} \\
\sigma_{yx} & \sigma_{yy} & \sigma_{yz} \\
\sigma_{zx} & \sigma_{zy} & \sigma_{zz}
\end{bmatrix}
\end{equation}
denotes the Cauchy stress tensor. The parameters used were
\begin{equation}
    \sigma_{\mathrm{GP}} = 0.02\,\mathrm{GPa},
    \qquad
    l = 1.5\,\upmu\mathrm{m},
\end{equation}
with mean stress vector
\begin{equation}
    \boldsymbol{\mu}_{\sigma}
    = (-0.005,\,-0.005,\,0.030,\,0,\,0,\,0)\,\mathrm{GPa}
\end{equation}
in Voigt notation
\((\sigma_{11},\sigma_{22},\sigma_{33},\sigma_{12},\sigma_{13},\sigma_{23})\). The stress field was sampled at the domain centres and at 40
additional boundary points around the sample, then linearly interpolated to the
tetrahedral centroids. The resulting stresses were converted to elastic strain
using the cubic aluminum stiffness constants
\begin{equation}
    C_{11}=104\,\mathrm{GPa},
    \qquad
    C_{12}=73\,\mathrm{GPa},
    \qquad
    C_{44}=32\,\mathrm{GPa},
\end{equation}
that is,
\begin{equation}
\mathbf{C}
=
\begin{bmatrix}
104 & 73  & 73  & 0  & 0  & 0 \\
73  & 104 & 73  & 0  & 0  & 0 \\
73  & 73  & 104 & 0  & 0  & 0 \\
0   & 0   & 0   & 32 & 0  & 0 \\
0   & 0   & 0   & 0  & 32 & 0 \\
0   & 0   & 0   & 0  & 0  & 32
\end{bmatrix}
\,\mathrm{GPa}.
\end{equation}
The compliance tensor \(\mathbf{S}=\mathbf{C}^{-1}\) was applied in the local
crystal frame of each tetrahedron, and the resulting strain tensor was
transformed back to the sample frame before being stored in the simulated
polycrystal.

The diffraction setup used a monochromatic X-ray beam with energy
\(50\,\mathrm{keV}\), corresponding to a wavelength of
\(0.247968\,\mathring{A}\). The beam propagated along the laboratory \(x\)-axis,
was polarized along \(y\), and had dimensions \(0.1\,\upmu\mathrm{m}\) in \(y\)
and \(0.22\,\upmu\mathrm{m}\) in \(z\) (covering the full height of the thin sample cylinder). Diffraction was recorded on a detector with \(2048\times2048\) pixels, a pixel size of
\(75\,\upmu\mathrm{m}\), and a sample-detector distance of
\(126.764\,\mathrm{mm}\). A scanning 3DXRD acquisition was simulated by
translating the sample in \(y\) with a step size of \(0.1\,\upmu\mathrm{m}\) over
103 positions. At each \(y\)-position the sample was rotated through
\(180^\circ\) in \(\omega\), using \(0.5^\circ\) rotation steps. Detector frames
were rendered with Lorentz, polarization, and structure-factor corrections as defined in \texttt{xrd\_simulator} \citep{henningsson2023}.

\begin{figure}[tbp]
\begin{center}
\includegraphics[width=0.5\textwidth]{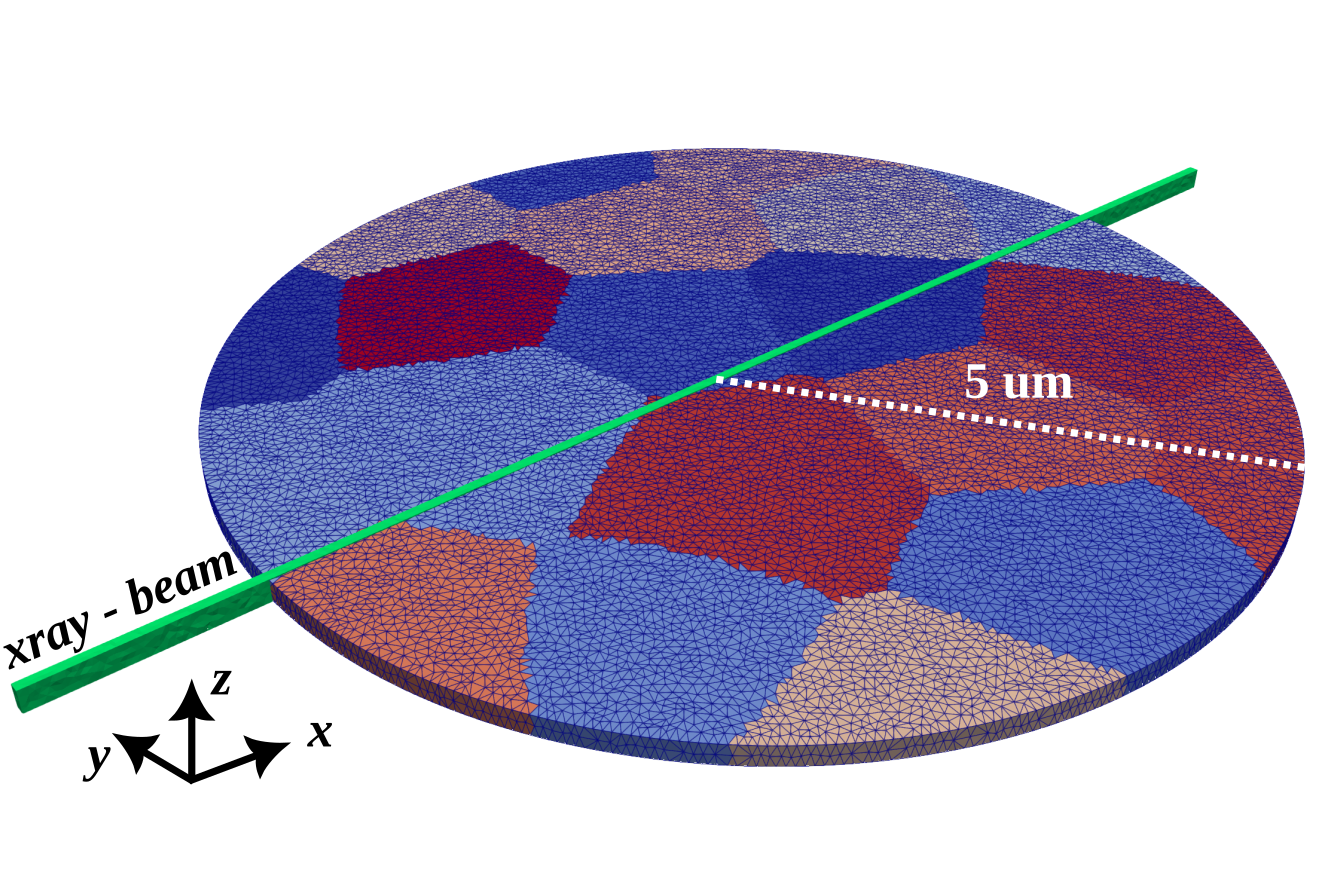}
\end{center}
\caption{Three-dimensional rendering of the simulated polycrystal. The thin cylindrical sample (radius $5\,\upmu\mathrm{m}$) is discretized into 163\,677 tetrahedral elements. The 20 grains are indicated by colour (grain index); the green line shows the X-ray beam direction along the laboratory $x$-axis.}
\label{simulation_3d}
\end{figure}

\begin{figure}[tbp]
\begin{center}
\includegraphics[width=0.8\textwidth]{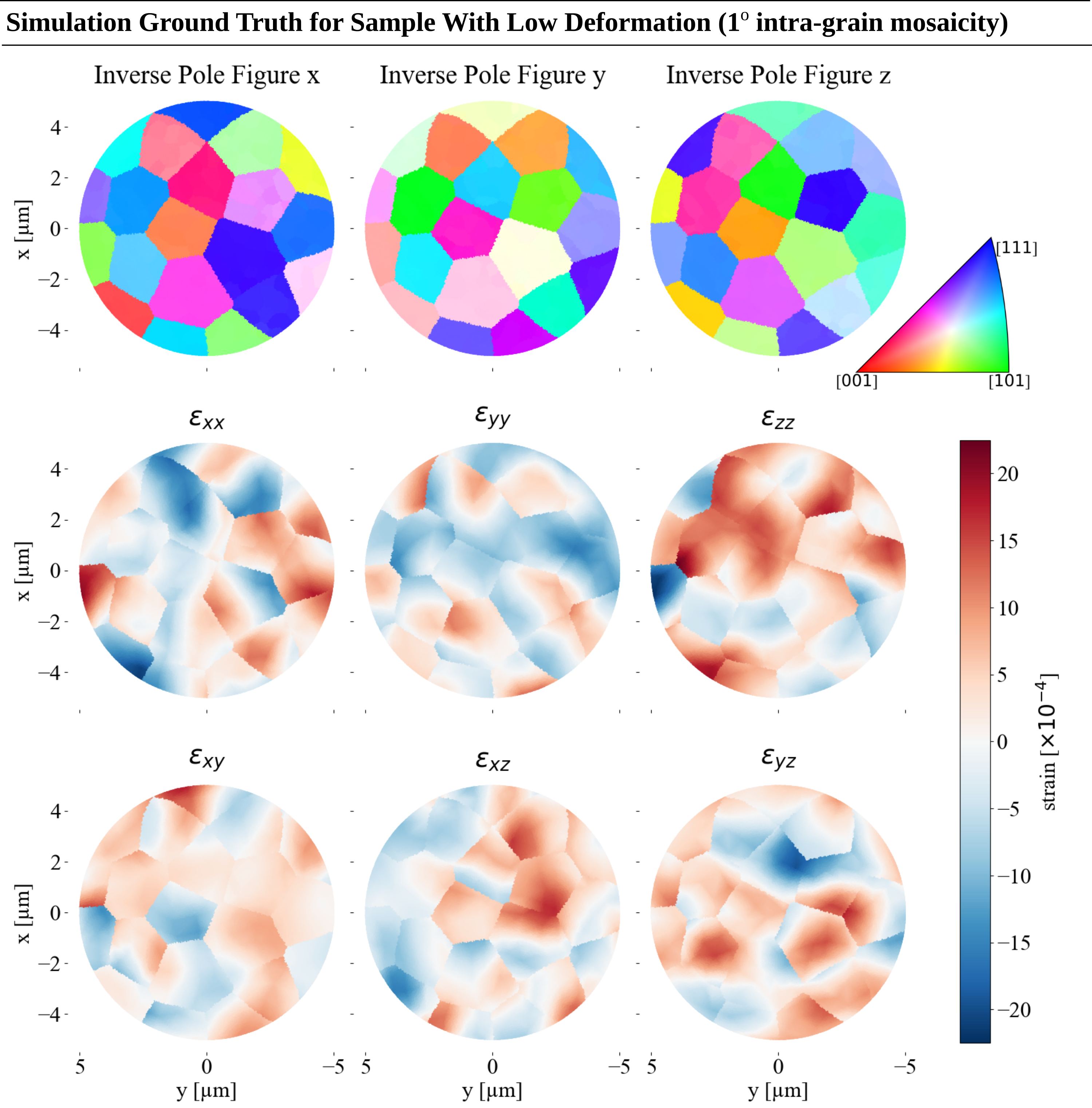}
\end{center}
\caption{Ground-truth for the $1^\circ$ mosaicity case. The figure displays the Inverse pole figure maps and components of the elastic strain tensor in the sample cross-section.}
\label{simulation_groundtruth_1deg}
\end{figure}

\begin{figure}[tbp]
\begin{center}
\includegraphics[width=0.8\textwidth]{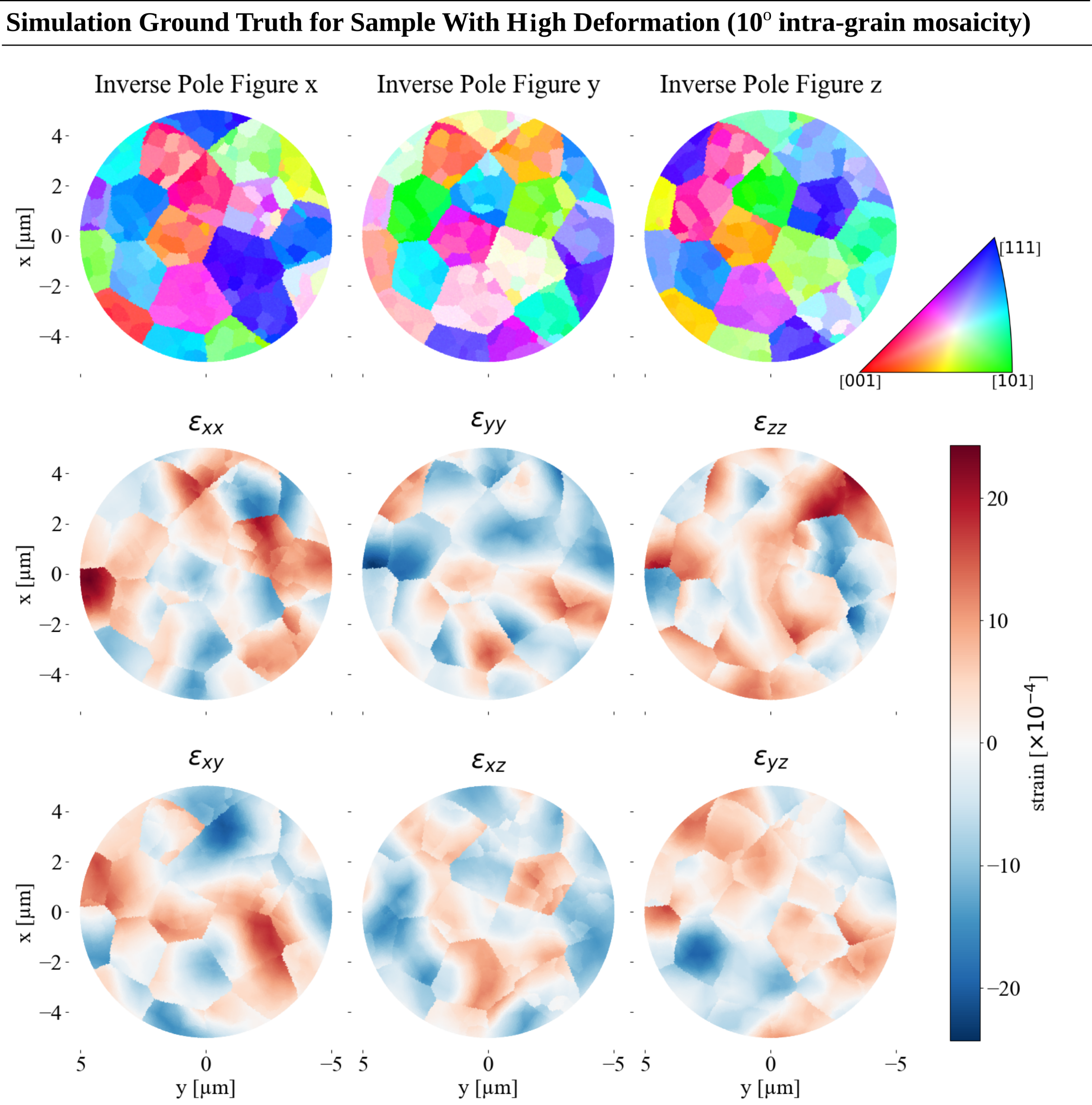}
\end{center}
\caption{Ground-truth fields for the $10^\circ$ mosaicity case. Layout as in Figure~\ref{simulation_groundtruth_1deg}. The larger mosaicity produces pronounced intragranular orientation gradients visible in the IPF map, while the imposed elastic strain field is identical in construction.}
\label{simulation_groundtruth_10deg}
\end{figure}

\section{Al1050 tensile deformation}\label{appendix:sample_deformation}

\begin{figure}[tbp] %
\begin{center}
\includegraphics[width=0.6\textwidth]{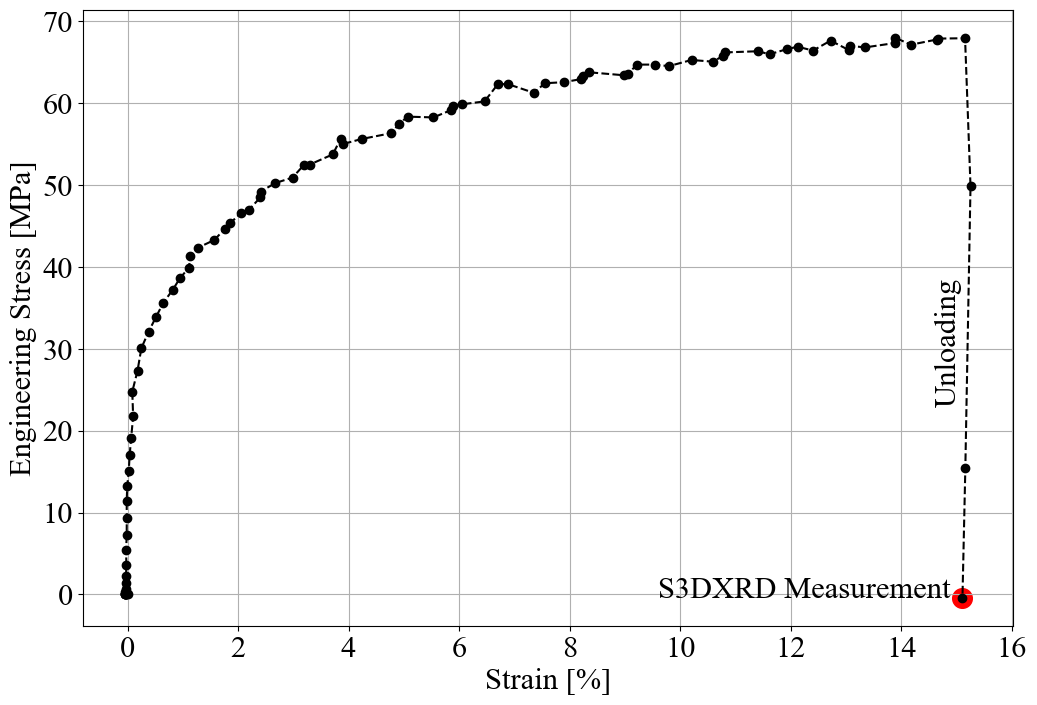}
\end{center}
\caption{Stress--strain curve for the Al1050 sample, showing a tensile elongation of $15\%$. The sample was cold rolled and heat treated prior to tensile deformation.}
\label{strain_stress_curve}
\end{figure}

Figure~\ref{strain_stress_curve} shows the engineering stress--strain curve recorded during the tensile test.

\section*{Acknowledgements}

Generative AI tools were used for language editing and for assistance with parts of the coding and manuscript preparation. The authors reviewed and validated all results and take full responsibility for the content of the article.

The authors thank Florencia Malamud for insightful discussions on the interpretation of the reconstructed texture distributions in the deformed specimen.

\section*{Funding}

The work of MSC was supported by Danish Data Science Academy, funded by the Novo Nordisk Foundation (NNF21SA0069429) and the Villum foundation (40516). We acknowledge financial support from the ERC Advanced Grant No. 885022.

The work of JSJ is supported by the Infrastructure for Quantitative AI-based Tomography (QUAITOM), which is supported by the Novo Nordisk Foundation (grant number NNF21OC0069766).

We acknowledge the MAX IV Laboratory for beamtime on the DanMAX beamline under proposal 20251117. Research conducted at MAX IV, a Swedish national user facility, is supported by Vetenskapsrådet (Swedish Research Council, VR) under contract 2018-07152, Vinnova (Swedish Governmental Agency for Innovation Systems) under contract 2018-04969 and Formas under contract 2019-02496. DanMAX is funded by the NUFI grant no. 4059-00009B.

\section*{Conflicts of Interest}

The authors declare no conflicts of interest.

\section*{Data Availability}

The aluminum specimen dataset will be made publicly available prior to publication. A compressed version of the simulated dataset is available with the Python library \citep{caroe2026}.

\bibliographystyle{abbrvnat}
\bibliography{references} % basename of .bib file

\end{document}